\documentstyle[aps,prd,eqsecnum]{revtex}
\input psfig

\begin{document}
\draft
\preprint{FERMILAB--Pub--97/xxx-A}
%\rightline{FERMILAB--Pub--97/xxx-A}
\preprint{astro-ph/9708???}
\preprint{submitted to {\it Physical Review D}}

\title{Constraints from High Redshift Supernovae upon Scalar Field Cosmologies}
\author{Joshua A.  Frieman}
\address{NASA/Fermilab Astrophysics Center,  
Fermi National Accelerator Laboratory\\   
PO Box 500, Batavia IL 60510, USA}   
\address{Department of Astronomy and Astrophysics\\
University of Chicago, Chicago, IL 60637 USA}
\author{Ioav Waga}
\address{Universidade Federal do Rio de Janeiro, Instituto de F\'\i sica\\ 
Rio de Janeiro, RJ, 21945-970, Brazil} 
\date{\today}
\maketitle
\begin{abstract}
Recent observations of high-redshift Type Ia supernovae have placed 
stringent constraints on the cosmological constant $\Lambda$.   
We explore the implications of these SNe observations 
for cosmological models in which a classically evolving scalar 
field currently dominates the energy density of the Universe. 
Such models have been shown to share the 
advantages of $\Lambda$ models: compatibility with 
the spatial flatness predicted inflation; a Universe  
older than the standard Einstein-de Sitter model; and, combined with cold dark 
matter, predictions for large-scale structure formation 
in good agreement with data from galaxy surveys.   
Compared to the cosmological 
constant, these scalar field models are consistent with 
the SNe observations for a lower matter density, $\Omega_{m0} \sim 
0.2$, and a higher age, $H_0 t_0 \gtrsim 1$.   
Combined with the fact that scalar field models imprint a distinctive 
signature on the cosmic microwave background anisotropy, they 
remain currently viable and should be testable in the near future. 

\end{abstract}

%\narrowtext

\section{Introduction}
\label{sec:level1}

In recent years, models with a relic cosmological constant $\Lambda$
have received considerable attention for a number of reasons. First,
dynamical estimates of the mass density on the scales of galaxy
clusters suggest that $\Omega_m  = 0.2 \pm 0.1$ for the matter $m$ that
clusters gravitationally (where $\Omega(t)$ is the ratio of the mean
mass density of the universe to the critical Einstein-de Sitter
density, $\Omega(t) = 8\pi G \rho/3H^2$) \cite{carlberg}.  (Some density
estimates on larger scales are higher but remain controversial
\cite{dbw97}.) However, if a sufficiently long epoch of inflation took
place during the early universe, the present spatial curvature should
be negligibly small, $\Omega_{tot} = 1$. A cosmological constant, with
effective density parameter $\Omega_\Lambda \equiv \Lambda/3H^2_0 = 1 -
\Omega_m$, is one way to resolve the discrepancy between $\Omega_m$ and
$\Omega_{tot}$.

The second motivation for the revival of the cosmological constant has
been the `age crisis' for spatially flat $\Omega_m = 1$ models, though
the evidence currently is more ambiguous than it was.  Estimates of the
Hubble expansion parameter from a variety of methods are converging to
$h \equiv (H_0/100~{\rm km/sec/Mpc})=0.7\pm 0.1$ \cite{hubble}, while
determinations of the age of the universe from globular clusters have
typically been in the range $t_{gc} \simeq 13 - 15$ Gyr or
higher\cite{gc}.  These observations imply a value for the `expansion
age', $H_0 t_0 = (H_0/ 70 ~{\rm km/sec/Mpc})(t_0/14 ~{\rm Gyr}) \simeq
1.0 \pm 0.2$, higher than that for the standard Einstein-de Sitter
model, for which $H_0 t_0 = 2/3$.  On the other hand, for models with a
cosmological constant, $H_0t_0$ can be larger: for example, for
$\Omega_\Lambda = 0.75 = 1-\Omega_m$, one finds $H_0 t_0 = 1.0$.  This
argument has recently been called into question, however: revised
determinations of $t_{gc}$ based on the Hipparcos distance scale are lower by
approximately 2 Gyr\cite{gc2}. If confirmed, this would largely
alleviate the age problem.

Third, cosmological constant-dominated models for large-scale structure
formation with cold dark matter (CDM) and a nearly scale-invariant
spectrum of primordial density perturbations (as predicted by
inflation) provide a better fit to the observed power spectrum of
galaxy clustering than does the `standard' $\Omega_m=1$ CDM model
\cite{lcdm}.  In particular, the shape of the power spectrum of galaxy
surveys is generally consistent with $\Gamma = \Omega_m h = 0.25 \pm
0.05$ \cite{pd}.  

Despite these successes, cosmological constant models face several
difficulties of their own.  On aesthetic grounds, it is difficult to
understand why the vacuum energy density of the universe, $\rho_\Lambda
\equiv \Lambda m^2_{Pl}/8\pi$, should be of order $(10^{-3} {\rm
eV})^4$, as it must be to have a cosmological impact ($\Omega_\Lambda
\sim 1$).  On dimensional grounds, one would expect it to be many
orders of magnitude larger -- of order $m_{Pl}^4$ or perhaps
$m_{SUSY}^4$.  Since this is not the case, we might plausibly assume
that some physical mechanism sets the ultimate vacuum energy to zero.
Why then is it not zero today?

In addition, the cosmological constant now faces strong observational
challenges. In $\Lambda$ models, a larger fraction of distant QSOs
would be gravitationally lensed than in a $\Lambda=0$ universe; surveys
for lensed QSOs have been used to infer the bound $\Omega_\Lambda <
0.66$ at 95 \% C.L.  \cite{kochanek}.  Further, while the power spectra of 
$\Lambda$
models with CDM have approximately the right shape to fit the galaxy
clustering data, the COBE-normalized amplitude is too high, requiring
galaxies to be anti-biased relative to the mass distribution~\cite{primack}.

Motivated by these theoretical and observational difficulties of the
cosmological constant, attention has turned to models in which the
energy density resides in a dynamical scalar field rather than in a
pure vacuum state. These {\it dynamical} $\Lambda$ models
\cite{dynlam,PR,RP,fhsw}\ were proposed in response to the aesthetic
difficulties of cosmological constant models. They were found to
partially alleviate their observational problems as well; for example,
the statistics of gravitationally lensed QSOs yields a less restrictive
upper bound on $H_0 t_0$ in these models\cite{fhsw,RQ,bw}.  In addition,
for a range of model parameters, the amplitude of the density power
spectrum is reduced relative to that of $\Lambda$CDM while its shape is
retained~\cite{cdf97}. These models also have a signature in the cosmic
microwave background (CMB) temperature anisotropy angular power
spectrum that is distinctive from $\Lambda$ and Einstein-de Sitter
models~\cite{cdf97,cds97}.  Consequently, they should be tested with
the next generation of high-resolution CMB temperature maps, e.g., from
the MAP and Planck satellite missions.

In this paper, we consider another set of observational constraints on
these cosmological models, arising from high-redshift
supernovae.  On-going projects to discover Type Ia supernovae at
redshifts $z \sim 0.3 - 1$, coupled with improved techniques to narrow
the dispersion in SN Ia peak magnitudes, have renewed the prospects for
determining the cosmological parameters \cite{perl96,schmidt97}.  Based
on analysis of an initial set of 7 high-redshift SNe Ia, Perlmutter
etal.  obtained the bound $\Omega_\Lambda < 0.51$ at 95\%
C.L.~\cite{perl96} for spatially flat cosmological constant models. 
For $\Lambda$ models, 
this implies $H_0 t_0 < 0.84$ at this limit. These are preliminary
results from a new method; the degree to which they are affected by
evolution, absorption, etc., will be determined by the much larger
samples now being gathered (the world sample of high-redshift SNe is
now roughly a factor of 10 larger than that used to obtain the bound
above).

We consider constraints on dynamical scalar field models arising from
high-redshift SNe Ia observations and compare them with constraints on
$\Lambda$ models.  The SNe Ia implications for some different but
related cosmological models--in which there is an extra component 
described by an
arbitrary fixed equation of state--have recently been studied in Refs.
\cite{vs}. Here, we focus on three representative models for
'ultra-light' scalar fields:  pseudo-Nambu-Goldstone bosons
(PNBGs)~\cite{fhsw}, inverse-power-law potentials~\cite{PR}, and
exponential potentials. In \S II, we review the motivation for and
cosmic evolution of these models.  In \S III, we derive the
corresponding constraints from the SNe observations. We conclude in \S
IV.

\section{Scalar Field Cosmological Models}
\label{sec:level3}

The classical action for a scalar field $\phi$ has the form
\begin{equation} 
S = \frac{m_{Pl}^2}{16 \pi}\int d^4x
\sqrt{-g}\left[\left(-R + \frac12 g^{\mu \nu} \partial_\mu\phi
\partial_\nu \phi - V(\phi)\right) + \mathcal{L}\right], 
\label{action}
\end{equation} 
where $m_{Pl} = G^{-1/2}$ is the Planck mass, $R$ is the
Ricci scalar, $g \equiv {\rm det} g_{\mu \nu}$, $V(\phi)$ is the scalar
field potential, and $\mathcal{L}$ is the Lagrangian density of
non-relativistic matter and radiation. For simplicity, we assume $\phi$
is minimally coupled to the curvature, and we work in units in which
$\hbar = c = 1$. We consider spatially flat, homogenous and isotropic
cosmologies described by the line element 
\begin{equation} 
ds^2= dt^2 -a^2(t) \left(dx^2 + dy^2 + dz^2\right)~, 
\end{equation} 
where $a(t)$ is
the Friedmann-Robertson-Walker (FRW) scale factor.

In this paper, we focus on models in which the scalar field is
dynamically important only at relatively recent epochs, at redshifts $z
\lesssim 10$. Thus, we model the matter content of the Universe as a
two-component system comprising the scalar field $\phi$ and
non-relativistic matter $m$.  Further, we assume that the
energy-momentum of each component is separately conserved, so the
matter energy density scales as $\rho_m \propto a^{-3}$.  (There is no
particle production as in some decaying $\Lambda$ models proposed in
the literature \cite{dynlam}.)

The Einstein and scalar field equations can be written as:
\begin{equation}
\frac{dH}{dt} + \frac32 H^2 + \frac{2 \pi}{m_{Pl}^2}\left(\frac{d\phi}{dt}
\right)^2 - \frac{4 \pi}{m_{Pl}^2} V(\phi) =0,
\label{dhdt}
\end{equation}
\begin{equation}
\frac{d^2\phi}{dt^2} + 3 H \frac{d\phi}{dt} + \frac{dV}{d\phi} = 0 
\label{dphidt}
\end{equation}
where the Hubble parameter
\begin{equation}
H^2 = \left(\frac1a\frac{da}{dt}\right)^2 = 
\frac{8\pi}{3m_{Pl}^2}\left[\frac12 \left({d\phi \over dt}\right)^2+ 
V(\phi) + \rho_m \right] ~.
\label{hubb}
\end{equation}
In what follows, it will also be 
useful to characterize the instantaneous equation of 
state of the scalar field by defining its effective adiabatic index, 
\begin{equation}
\gamma_\phi(t) = 1 + \frac{p_\phi}{\rho_\phi} = {2\left({d\phi \over dt}
\right)^2 \over \left({d\phi \over dt}\right)^2 + 2V(\phi)}~.
\label{gamdef}
\end{equation}
For a static field, corresponding to a cosmological constant $\Lambda$, 
$\gamma_\Lambda =0$, while for pressureless dust, $\gamma = 1$.

A number of models with a dynamical $\Lambda$ have been discussed in
the literature \cite{dynlam,PR,RP,fhsw}. 
We consider three representative scalar
field potentials that give rise to effective decaying $\Lambda$
models.

\subsection{The PNGB model}

Consider the properties that a massive scalar field must satisfy in
order to act approximately like a cosmological constant at recent epochs. 
Vacuum energy is stored in the potential energy density $V(\phi) \sim M^4$, 
where $M$ sets the characteristic height of the
potential, and we set $V(\phi_m)=0$ at the minimum of the potential by
the assumption that the fundamental vacuum energy of the Universe is
zero (for reasons not yet understood).  In order to generate a non-zero
$\Lambda$ at the present epoch, $\phi$ must initially be displaced from
the minimum ($\phi_i \neq \phi_m$ as an initial condition), and its
kinetic energy must be relatively small compared to its potential
energy.  This implies that the motion of the field is still (nearly)
overdamped, so the scalar mass must be extremely small, $m_\phi \equiv
\sqrt{|V''(\phi_i)|} \lesssim 3H_0 = 5\times 10^{-33} h$ eV. In addition,
for $\Omega_\phi \sim 1$, the potential energy density should be of
order the critical density, $M^4 \sim 3H^2_0 m^2_{Pl}/8\pi$, or $M
\simeq 3\times 10^{-3}h^{1/2}$ eV.  Thus, the characteristic height and
curvature of the potential are strongly constrained for a classical
model of the cosmological constant.

In quantum field theory, such ultra-low-mass scalars are not {\it
generically} natural: radiative corrections generate large mass
renormalizations at each order of perturbation theory. To incorporate
ultra-light scalars into particle physics, their small masses should be
at least `technically' natural, that is, protected by symmetries, such
that when the small masses are set to zero, they cannot be generated in
any order of perturbation theory, owing to the restrictive symmetry.

From the viewpoint of quantum field theory, pseudo-Nambu-Goldstone
bosons (PNGBs) are the simplest way to have naturally ultra--low mass,
spin--$0$ particles.  PNGB models are characterized by two mass scales,
a spontaneous symmetry breaking scale $f$ (at which the effective
Lagrangian still retains the symmetry) and an explicit breaking scale
$M$ (at which the effective Lagrangian contains the explicit symmetry
breaking term). The PNGB mass is then $m_\phi \sim M^2/f$. Thus, the
two dynamical conditions on $m_\phi$ and $M$ above essentially fix
these two mass scales to be $M \sim 10^{-3}$ eV, interestingly close to
the neutrino mass scale for the MSW solution to the solar neutrino
problem, and $f \sim m_{Pl} \simeq 10^{19}$ GeV, the Planck scale.
Since these scales have a plausible origin in particle physics models,
we may have an explanation for the `coincidence' that the vacuum energy
is dynamically important at the present epoch \cite{fhsw,fhw,fukyan}.
Moreover, the small mass $m_\phi$ is technically natural.

The effective scalar field potential in PNGB models is approximated by 
\begin{equation}
V(\phi) = M^4 \left(1 + \cos\left(\phi / f\right)\right) ~.
\label{pngbV} 
\end{equation}
Constraints on the $f-M$ parameter space
from gravitational lensing were analyzed in \cite{fhsw}; the
large-scale power spectrum and CMB anistropy for these models were
studied in \cite{cdf97,cds97,vl97}.

To numerically integrate the field equations, we 
define dimensionless variables, 
\begin{eqnarray} 
u = \frac{1}{H_0
f}\frac{d\phi}{dt} ,\; v = \frac{H}{H_0} \;\;\ \mbox{and} \;\;\; w =
\frac{\phi}{f}.  
%\label{dimvar}
\end{eqnarray} 

\begin{figure} 
\hspace*{1.3in}
\psfig{file=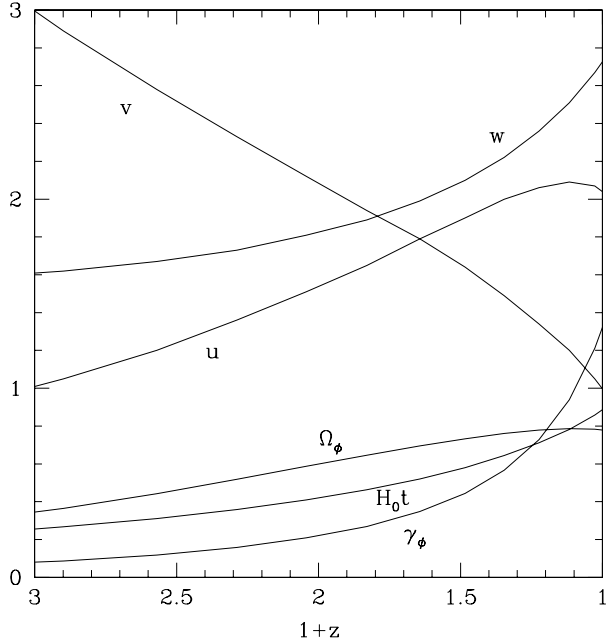,height=9.cm,width= 9.cm} 
\vspace*{0.4in}
\caption{The quantities $u, v, w$, $\Omega_{\phi}$, $H_0t$, and $\gamma_\phi$ 
as a function of
redshift for the PNGB model with $f=2.07\times 10^{18}$ GeV and $M
=0.004 h^{1/2}$ eV. The initial field conditions are chosen to be 
$w_i=\phi_i/f=1.5$, $u_i=0$.} 
\end{figure}

\noindent With these definitions, we can rewrite the equations of motion in 
first order form,
\begin{equation} 
\dot{w} = u, 
\end{equation} 
\begin{equation} 
\dot{u} + 3 u v - \frac{M^4}{H_0^2f^2} \sin w = 0, 
\end{equation}  
\begin{equation} 
\dot{v} +
\frac32 v^2 + \frac{2\pi f^2}{m_{Pl}^2} u^2 - \frac{4\pi M^4}{m_{Pl}^2H_0^2}
\left(1+\cos w\right)=0.  
\end{equation} 
Here the overdot denotes $(1/H_0)d/dt$.  We numerically solve the above
equations assuming that $u(t_{i})=0$ and that $v(t_{i})\gg 1$ (so the 
universe is initially matter-dominated).

In Fig. 1 we show the quantities $ u, v, w$, $\Omega_{\phi}$, 
$H_0t$, and $\gamma_\phi$ 
 as a function of redshift, $1+z=a_0/a(t)$, for the parameters
$f=2.07\times 10^{18}$ GeV and $M=0.004 h^{1/2}$ eV. The initial
conditions for the field were taken to be $w_i=\phi_i/f=1.5$, $u_i=0$.
For this choice of parameters and initial conditions, 
$\Omega_{\phi 0}=0.78$ and $H_0 t_0 = 0.89$. 
(For comparison, for a $\Lambda$ model with $\Omega_\Lambda=0.78$, 
we would have $H_0 t_0 = 1.05$; in an open model with the 
same value of $\Omega_{m0}=0.22$, the corresponding age is 
$H_0 t_0 = 0.84$.) The $u$ and $w$ curves 
indicate that the field is just beginning to decelerate at 
redshift $z \simeq 0.1$, as it nears the potential minimum ($w_m=\pi$) for the 
first time. At high redshift, when the field is nearly static, 
the adiabatic index $\gamma_\phi \simeq 0$, and the field acts as 
a pure cosmological constant; at late times, when the field kinetic 
energy becomes appreciable, $\gamma$ rises above unity. In the future, 
the field would undergo damped 
oscillations around the minimum at $w=\pi$, and $\gamma$ 
would settle down to unity, the value for pressureless dust.

In Figs. 2 and 3, we plot contours of $\Omega_{\phi 0}$ and $H_0 t_0$
as a function of the parameters $f$  and $M$, also for
$w_i=\phi_i/f=1.5$.  (For different choices of $w_i$, the contour
levels would shift around in the $f-M$ plane; for comparison, see
~\cite{fhsw,cdf97}.) These figures show that there is a range of 
model parameters which give rise to acceptable values of $\Omega_{\phi 0}$ 
and $H_0 t_0$. We also note that, in the region of parameter space studied 
here, the linear transfer function for the growth of large-scale structure 
has an effective shape parameter given by $\Gamma = (1-\Omega_{\phi 0})
h$, but the perturbation amplitude can differ from that in the 
corresponding $\Lambda$ model~\cite{cdf97}.

\begin{figure} 
\hspace*{1.3in}
\psfig{file=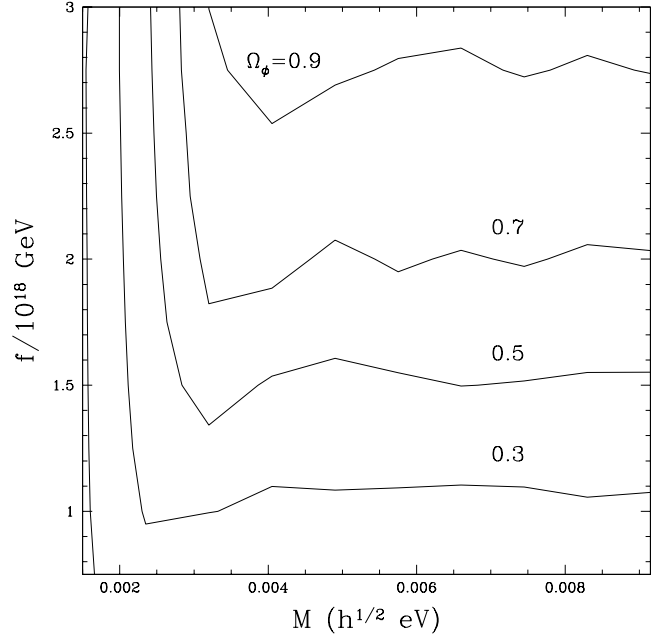,height=9.cm,width= 9.cm}
\vspace*{0.4in}
\caption{Contours of constant $\Omega_{\phi 0}$ in the $f-M$ plane 
for the PNGB model, with $w_i=1.5$.}
\end{figure}

\begin{figure} 
\hspace*{1.3in}
\psfig{file=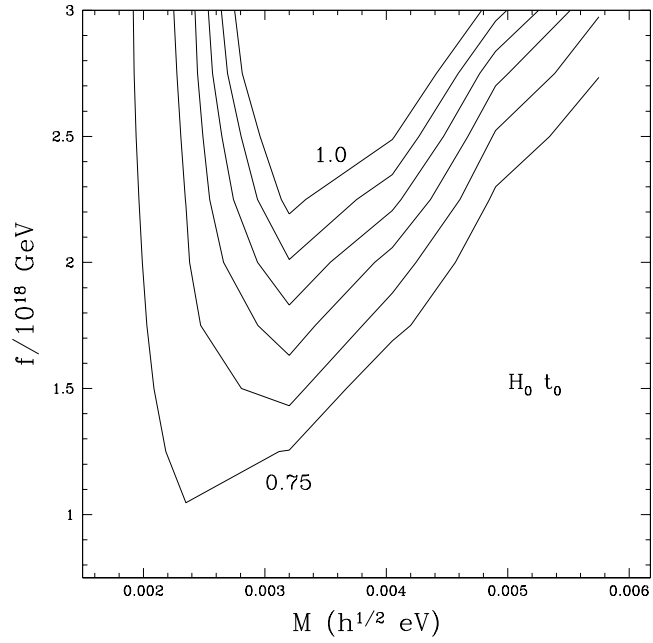,height=9.cm,width= 9.cm}
\vspace*{0.4in}
\caption{Contours of constant $H_0 t_0$ in the $f-M$ plane 
for the PNGB model, with $w_i=1.5$.}
\end{figure}

\subsection{Power-law potentials}

For these models the scalar field potential has the form of an inverse 
power-law,
\begin{equation}
V(\phi) = \frac{k}{32\pi}m_{Pl}^4 \left({m_{Pl}\over \sqrt{16\pi}\phi}
\right)^{\alpha}~,
\label{Vpl}
\end{equation}
where  $k > 0$ and $\alpha > 0$ are dimensionless constants. 
Scalar potentials of this form arise, e.g., in dynamically broken 
supersymmetry theories in which flat directions are lifted by non-perturbative 
effects~\cite{kr97}. However, for such a field to be dynamically 
relevant today requires $k \sim 10^{-120}$; this is just another 
statement of the cosmological constant problem. 

Cosmological consequences of scalar fields with such a 
potential were investigated in \cite{PR,RP,RQ}. 
For $\alpha \rightarrow 0$, the scalar field energy-momentum tensor 
approaches that of a 
conventional cosmological constant $\Lambda$, i.e., $\rho _{\phi} =$ constant; 
in the opposite limit $\alpha \rightarrow \infty$, the scalar field energy 
density scales like that of non-relativistic matter, 
$\rho_{\phi} \propto a^{-3}$. More generally, in the matter-dominated era 
at $z \gg 1$, when $\rho_\phi \ll \rho_m$, the scalar field energy density 
scales as $\rho_{\phi} \propto a^{-3 \alpha/(\alpha+2)}$. 
Thus, for fixed $\Omega_{\phi 0}$, 
the angular diameter distance to a fixed redshift, and thus the 
optical depth for gravitational lensing,  
decreases as $\alpha$ increases. Unlike the case of a cosmological 
constant, in these models 
it is possible to satisfy the lensing
constraints~\cite{kochanek} even for low values of 
$\Omega_{m0}$ (see Ref.\onlinecite{RQ}). 
As we shall see in the next section, similar statements apply for the high
redshift supernovae constraints.

By defining dimensionless variables,
\begin{eqnarray}
u = \frac{4\sqrt{\pi}}{H_0 m_{Pl}}\frac{d\phi}{dt} ,\; 
v = \frac{H}{H_0}, \; \mbox{and} \; w = \frac{4\sqrt{\pi}\phi}{m_{Pl}}.
\end{eqnarray}
the field equations can be rewritten as:
\begin{equation}
\dot{w} = u,
\end{equation}
\begin{equation}
\dot{u} + 3 u v - \frac{\alpha}{2}\frac{k m_{Pl}^2}{H_0^2} w^{-(\alpha +1)} = 0,
\end{equation}
\begin{equation}
\dot{v} + \frac32 v^2 + \frac{u^2}{8} - \frac18 \frac{k m_{Pl}^2}{H_0^2} 
w^{-\alpha}=0.
\end{equation}
We numerically evolve the fields using the initial conditions
$u(t_i) = 0$, $v(t_i) \gg 1$ as before. 
For fixed values of the model parameters $\alpha$ and $k$, the choice of 
the initial field value $w(t_i)$ determines the cosmological parameters 
$\Omega_{\phi 0}$ and $H_0 t_0$. Alternatively, we can keep $\alpha$ and 
$w(t_i)$ fixed and use $\Omega_{m0}=1-\Omega_{\phi 0}$ 
rather than $k$ as our free 
parameter; we shall follow this approach below. 

As an example, Fig. 4 
shows the evolution for $\alpha=5$, $w(t_i)=3$, and $\Omega_{m0}=0.2$.
For this model $H_0 t_0 = 0.92$, larger than the value $H_0  t_0 = 0.85$ 
obtained in an open model with the same value of $\Omega_{m0}$. 
In the next section, we shall see that this choice of parameters and 
initial conditions is consistent with the high-$z$ SNe Ia constraints. 

\begin{figure} 
\hspace*{1.3in}
\psfig{file=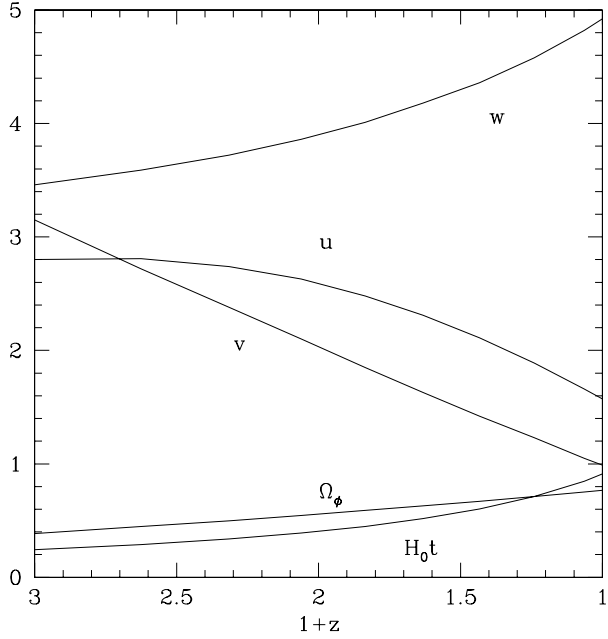,height=9.cm,width= 9.cm}
\vspace*{0.5in}
\caption{Evolution of the variables 
$u, v, w$, $\Omega_{\phi}$, and $H_0 t_0$ for a power-law potential
with $\alpha=5$, $w(t_i)=3$, and $\Omega_{m0}=0.2$.}
\end{figure}

\subsection{Exponential potentials}

In this case the scalar field potential has the functional form,
\begin{equation}
V(\phi) = V_0 \;e^{- \phi/f},
\end{equation}
with positive constants $V_0$ and $f$. 
Scalar fields with an exponential potential have been investigated in the 
context of power-law inflationary models \cite{lucch}. 
Cosmological consequences of scalar fields with 
exponential potentials dominating the dynamics 
of the Universe at late times were analyzed in \cite{RP,sahni92}.  

Again introducing dimensionless variables,
\begin{eqnarray}
u = \frac{1}{H_0 f}\frac{d\phi}{dt} ,\; v = \frac{H}{H_0} \;\;\;\;
\mbox{and} \;\;\;\; w = \frac{\phi}{f} - \ln \left(\frac{V_0}{m_{Pl}^2 H_0^2}
\right), 
\label{expdimvar}
\end{eqnarray}
the field equations become 
\begin{equation}
\dot{w} = u,
\end{equation}
\begin{equation}
\dot{u} + 3 u v - 8 \pi \beta \;e^{-w}  = 0,
\end{equation}
\begin{equation}
\dot{v} + \frac32 v^2 + \frac{u^2}{4 \beta} - 4 \pi  e^{-w} =0,
\end{equation}
where $\beta = m_{Pl}^2/8 \pi f^2$. We numerically 
evolve the fields with the initial conditions 
$u(t_i) = 0$, $v(t_i) \gg 1$. The mass parameter $\beta$ and the initial 
field value $w(t_i)$ determine the cosmological 
parameters $\Omega_{m0}$ and $H_0  t_0$. 
We note that $V_0$ is not a fundamental constant: as Eq.(\ref{expdimvar}) shows, 
changing $V_0$ is equivalent to reescaling the scalar field  $\phi$.  

\begin{figure} 
\hspace*{1.3in}
\psfig{file=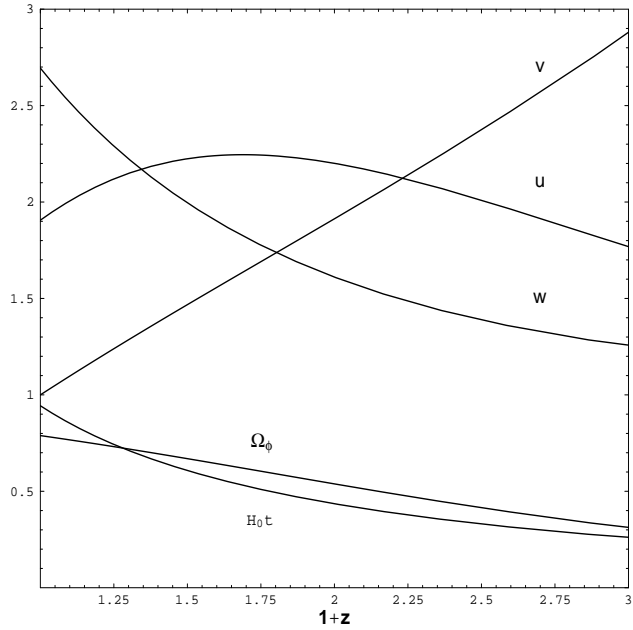,height=8.2cm,width= 8.2cm}
\vspace*{0.5in}
\caption{Evolution of the variables 
$u, v, w$, $\Omega_{\phi}$, and $H_0 t$ with redshift 
in the exponential potential model 
with $w(t_i)=1$ and $\ln \beta =1$.}
\end{figure}

In Fig. 5, we show the evolution of the quantities $u$, $v$, $w$, 
$\Omega_{\phi}(t)$, and $H_0 t$ with redshift $z$ for the parameter 
choice $\beta = 2.72$ and the initial condition $w(t_i)=1$.
For this case, we obtain $\Omega_{m0}=0.21$ and $H_0  t_0 = 0.94$; by 
comparison, in an 
open model with the same value of $\Omega_{m0}$, we would have 
$H_0  t_0 = 0.84$.
In the next section we show that this choice of 
model parameters is consistent with the SNe Ia data.  

\begin{figure} 
\hspace*{1.3in}
\psfig{file=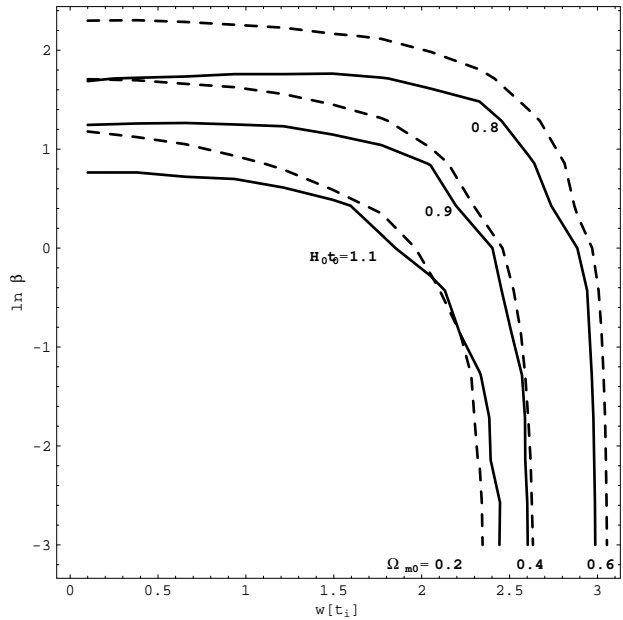,height=8.2cm,width= 8.2cm}
\vspace*{0.3in}
\caption{Contours of constant $\Omega_{m0}$(dashed curves) and $H_0 t_0$ 
(solid) in the 
$\ln \beta - w(t_i)$ parameter space for exponential potentials.}
\end{figure}

In Fig.$6$ we show contours of constant $\Omega_{m0}$ (dashed curves) and 
$H_0 t_0$ (solid) in the $\ln{\beta} - w(t_i)$ parameter space. 
From the point of view of both large scale structure 
and age constraints, the 
most interesting region of the parameter space would
seem to be the bottom right-hand portion
of the figure, the locus of highest $H_0 t_0$ 
for fixed $\Omega_{mo}$.  However, as we shall see in the next section, 
the SNe Ia 
constraints practically exclude this region. We will show that for  
values of the mass parameter
$\ln \beta \lesssim -2$, the SNe constraints imply $w(t_i)\gtrsim 3$;  
for these values of 
the parameters, however, $\Omega_{m0} \gtrsim 0.6$ and $H_0 t_0 \lesssim 0.8$ .

\section{Constraints from high-redshift Type Ia Supernovae}

\subsection{ The SNe observations} 

There are now two major 
ongoing programs to systematically discover high-redshift 
supernovae. In a recent report Perlmutter {\it et
al.} \cite{perl96} analyzed a first set of 
seven Type Ia SNe with redshifts $z=0.35 -
0.46$ and obtained constraints on the cosmological parameters.
Their preliminary result, $\Omega_{\Lambda} < 0.51$ at the $95\%$
confidence level, strongly constrains models with a
cosmological constant. In this section we use these data to constrain 
the scalar field cosmological models described in the
preceding section.

The essential idea behind the technique is to apply the classical 
redshift-magnitude test to SNe Ia as standard candles. For a 
source of absolute magnitude $M$, the
apparent bolometric magnitude $m(z)$ can be expressed as 
\begin{equation} m(z) =
{\cal{M}} + 5 \;\log d_l, 
\label{appmag}
\end{equation} 
where $d_l$ is the luminosity
distance in units of $H_0^{-1}$, and 
\begin{equation} {\cal{M}} =
\mbox{M} - 5 \;\log H_0 + 25 
\end{equation} 
is the ``zero point''
magnitude (or Hubble intercept magnitude), estimated from observations 
of low-redshift ($z < 0.1$) SNe Ia. The
nearby supernovae data set used in \cite{perl96} to determine 
$\cal{M}$ comprised those 18 SNe Ia discovered in the Calan/Tololo
Supernovae Search \cite{hamuy} for which the first observations were
made no later than 5 days after maximum.

Arising from the explosion of accreting white dwarfs, SNe Ia 
do not constitute a completely homogeneous class: there is 
significant dispersion in their absolute magnitudes at maximum light.  
However, it has been shown that SN Ia peak absolute magnitude 
is correlated with the rate at which the light curve subsequently 
declines \cite{phil93}: brighter SNe Ia fade more slowly. 
The rate of decline can be quantified, e.g., by 
$\Delta m_{15}$, the B-magnitude decline in the first 15 days after
maximum. For the Calan/Tololo sample, correction of the observed 
$B$-magnitudes using $\Delta m_{15}$ reduced the dispersion in 
peak absolute magnitude from $\sigma_{M_B,corr} = 0.26$ to 0.17. A similar 
procedure applied to the Perlmutter {\it et al.} sample achieved 
comparable results, reducing $\sigma_M$ from 0.27 to 0.19 mag. 
The width-luminosity correlation has been developed with the 
light-curve shape method \cite{rpk95} and further refined with the 
use of multiple pass bands \cite{rpk96}.

In our computations we follow \cite{perl96} and use the corrected
B-magnitude intercept at $\Delta m_{15}=1.1$ mag,
${\cal{M}}_{B,corr}^{\{1.1\}}= -3.32 \pm 0.05$. Of the 7 SNe Ia in 
the high-redshift sample, we consider only those
5 that satisfy $0.8 < \Delta m_{15} < 1.5$, corresponding to  
the range of values covered by the calibrating set of 18 low-redshift 
supernovae. To construct the $\chi^2$ values, we used the 
outer error bars of the Ref.\cite{perl96} data points, obtained by adding in
quadrature the error bars of $m_{B,corr}$ (the apparent B-magnitudes
after width-luminosity correction) to 
$\sigma_{M_B,corr}$.

\subsection{Results}

\subsubsection{The PNGB model}

We calculate the apparent magnitude-redshift relation for a grid of 
PNGB models in the $f-M$ parameter space and compare with the 
high-redshift SNe Ia observations.  
In Fig. $7$ we show the corresponding $95\%$, $90\%$, and $68\%$
confidence level bounds on the parameters $f$ and $M$. As for Figs. 2 
and 3, these limits apply to models with the initial condition 
$w(t_i)=1.5$; for other choices, the bounding contours would shift 
by small amounts in the $f-M$ plane.  

\begin{figure}
\hspace*{1.3in} 
\psfig{file=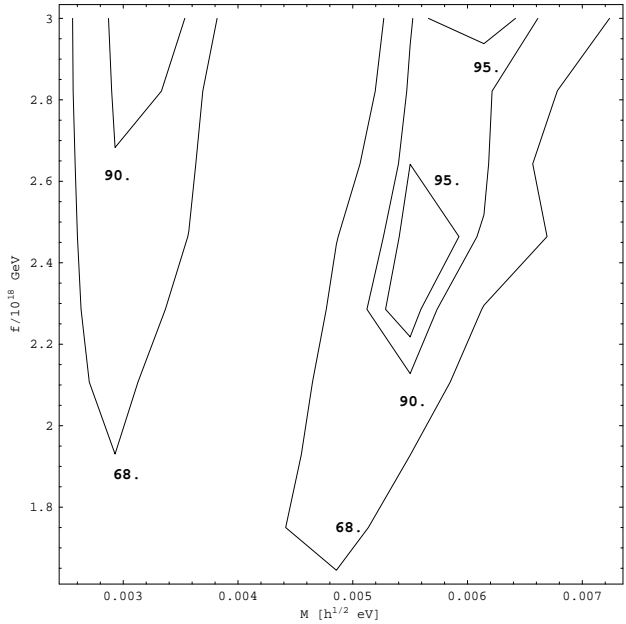,height=8.2cm,width= 8.2cm}
\vspace*{0.2in} 
\caption{Limits on the $f-M$ parameter space of PNGB models from 
the first set of high-redshift SNe Ia, for $w(t_i)=1.5$; the 
lowest and highest contours are $1-$ and $2-\sigma$ limits.}
\end{figure}

Note the existence of two excluded
regions of parameter space, one at the left and the other 
in the right portion of the
figure. To understand the meaning of these regions, 
consider three cases with fixed $f=3\times 10^{18}$ GeV and varying  
$M=0.003$, $0.0045$, and $0.006$ eV. The first and third choices 
are excluded by the SNe data while the second is allowed. With 
increasing $M$, the corresponding values of 
$\Omega_{\phi 0}$ and $H_0 t_0$ are 
(0.83, 1.07), (0.96, 1.11), and (0.80,
0.93). In all three cases, the Universe is $\phi$-dominated ($\Omega_{\phi} 
\gg \Omega_{m}$) for $z \lesssim 1$, but the evolution differs markedly 
between them. To see this, 
in Fig. $8$ we show the effective adiabatic index of the scalar
field, $ \gamma_\phi(t)$, as a function of
redshift for the three cases.

\begin{figure} 
\hspace*{1.3in}
\psfig{file=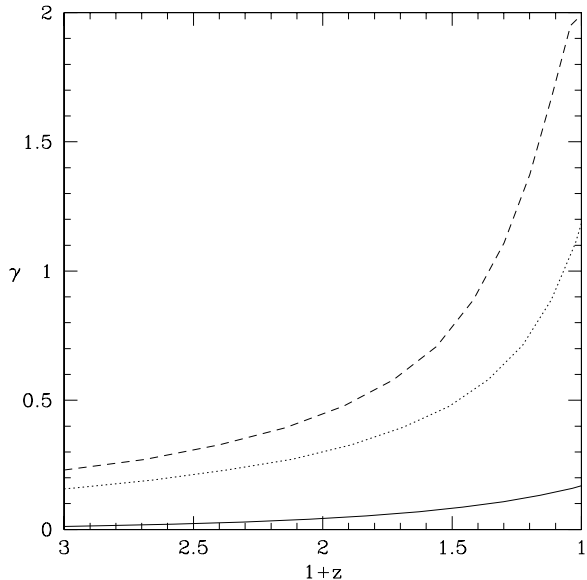,height=8.2cm,width= 8.2cm}
\vspace*{0.5in}
\caption{Evolution of the equation of state parameter $\gamma_\phi$ 
with redshift for three PNGB models with $f=3\times 10^{18}$ GeV and 
$w(t_i)=1.5$: $M=0.003$ eV (solid), $M=0.0045$ (dotted), and 
$M=0.006$ eV (dashed).}
\end{figure}

For the first case, $M=0.003$ eV, 
$\gamma_\phi$ remains close to zero throughout 
the evolution; in this case, the 
low value of $M$ implies that the effective scalar mass $m_\phi \lesssim 
3H_0$, and the nearly static scalar field behaves approximately 
like a cosmological constant until the present epoch. As a 
result, the redshift-magnitude relation for this case will be 
similar to that of a $\Lambda$ model with $\Omega_\Lambda \simeq 0.83$, 
which is excluded by the SNe Ia data. 
In the second case, $M=0.0045$ eV, the evolution of $\gamma_\phi$ is 
more pronounced, increasing from $\gamma \sim
0.3$ at $z=1$ to $\gamma \sim 1.2$ at $z=0$. At the moderate 
redshifts probed by the current SNe observations, 
$z \lesssim 0.4$, the effective 
equation of state in this case does not differ drastically from 
that of the Einstein-de Sitter model ($\gamma=1$), which is consistent with 
the SNe data. In the third case, with 
$M=0.006$ eV, $\gamma$ increases to large values at recent epochs, 
again producing a distance-redshift relation appreciably different 
from that of the Einstein-de Sitter model.

\begin{figure}
\hspace*{1.3in} 
\psfig{file=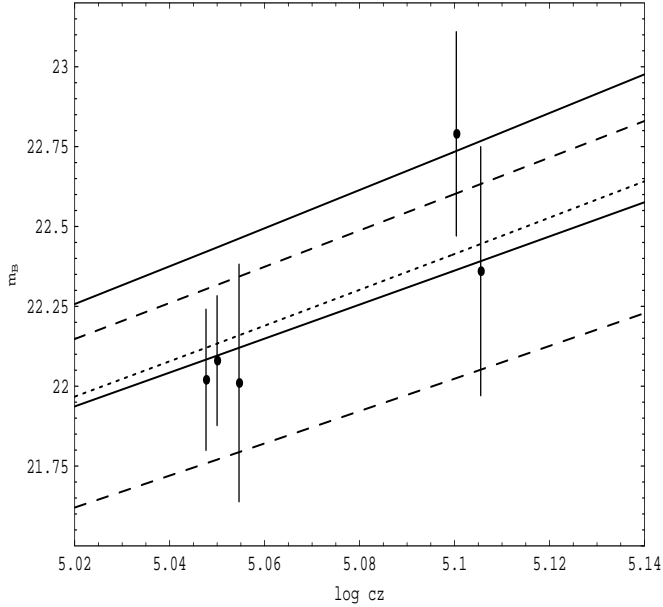,height=8.2cm,width=8.6cm}
\vspace*{0.2in} 
\caption{Apparent magnitude vs. redshift relation is shown for the 
3 PNGB models corresponding to Fig.8: 
$M=0.003$ eV (top solid line); 0.0045 eV (middle dotted line); 0.006 
eV (bottom dashed line). For comparison, we also show the 
prediction for the standard $\Omega_m=1$ Einstein-de Sitter model 
(middle solid line), and for the $\Lambda$ model at the 95 \% C.L. 
limit, $\Omega_\Lambda=0.51$ (top short dashed line).
The data points are the light 
curve-corrected data for the 5 high-$z$ SNe Ia.} 
\end{figure}

In Fig. $9$
we display the apparent magnitude-redshift relation 
for these three cases along with  the
corrected magnitudes for the five high-redshift SNe Ia 
used in this analysis.
For comparison we also show the prediction of the Einstein-de
Sitter model and the $\Lambda$ model at the 95\% C.L. limit.
The $M=0.003$ eV case is ruled out because, as in  
the $\Lambda$ model, SNe at fixed redshift  
should be brighter than observed; in the $M=0.006$ eV case, sources 
are too faint. 

Thus, the behavior of the effective scalar equation of state provides a 
qualitative understanding of the topology of the exclusion regions in 
Fig. 7. By comparing Fig. $7$ with Figs. $2$ and $3$, we see that the allowed 
region of parameter space includes models with low $\Omega_{m0}$ and a 
relatively high value of $H_0 t_0$ (as compared with open models with the same
$\Omega_{m0}$). For example, for 
$f=2.0 \times 10^{18}$ GeV and $M = 0.0035$ eV, parameter values 
consistent with the SNe data, we have 
$\Omega_{m0}=0.25$ and $H_0 t_0 = 0.92$. An open model with the same value of 
$\Omega_{m0}$ would correspond to $H_0 t_0 = 0.83$. 
A particular interesting region of parameter space is the area 
around $f=1.8 \times 10^{27}$ eV and $M =
0.003$ eV, in the protuberance of Fig. 3. For these parameter values, the 
age is relatively high, $H_0 t_0 =0.87$, the magnitude-redshift relation 
is consistent with the SNe data, and the present matter density is 
$\Omega_{m0}=0.38$. With CDM and normalized to COBE, this model predicts 
a large-scale power spectrum consistent with the galaxy clustering data 
as well \cite{cdf97}. 

\subsubsection{Power-Law Potentials}

As noted above, for fixed $w(t_i)$, 
the model parameters for the power-law potentials 
can be taken to be $\Omega_{m0}$ and $\alpha$.
In Fig.$10$ we show the $95\%$, $90\%$, and $68\%$ C. L. limits from 
the SNe Ia data on the parameter space for these models. As in 
Fig. 4, we have fixed $w(t_i)=3$. We also display the 
contours of constant $H_0 t_0$. For 
$\Omega_{m0} = 0.2$, 0.3, and 0.4, the $1-\sigma$ SNe limits are 
$\alpha \geq 4.45$, $4.07$, 
and $3.6$, respectively, and the corresponding upper limits on 
$H_0 t_0$ are 0.95, 0.91, and 0.86. 

\begin{figure}
\hspace*{1.3in} 
\psfig{file=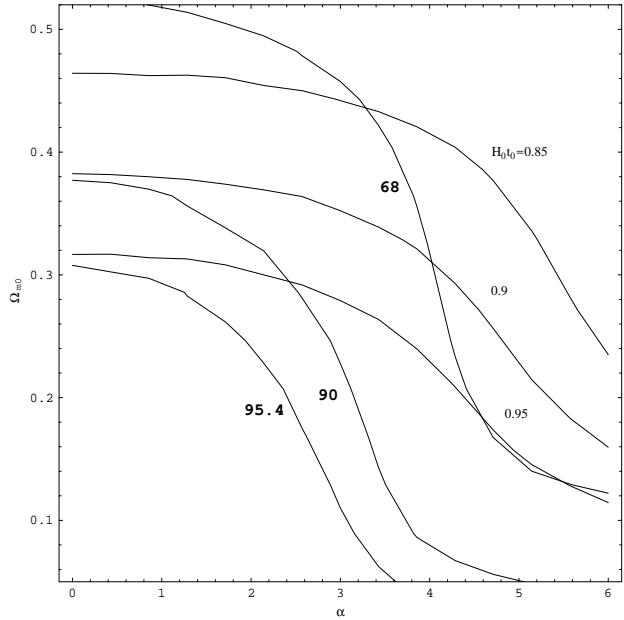,height=8.2cm,width= 8.2cm}
\vspace*{0.1in} 
\caption{Contours for the $1-\sigma$, $90\%$ C.L., and $2-\sigma$ 
SNe Ia limits 
in the $\alpha - \Omega_{m0}$ parameter space for power-law 
potentials. Also shown are  
contours of constant
$H_0 t_0 = 0.85$, 0.9, and 0.95.}
\end{figure}

\subsubsection{Exponential Potentials} 

In Fig.$11$ we show the 
$95.4\%$, $90\%$, and $68\%$ C. L. SNe Ia limits  on the 
$\ln \beta - w(t_i)$ parameter space for the exponential potential models.
As noted in the discussion of Fig. 6, the region in the
bottom right portion of the figure is not cosmologically interesting: 
for $w(t_i) \gtrsim 2.9$, as required at $1-\sigma$ by the 
SNe data, Fig. 6 indicates that 
$\Omega_{m0} \gtrsim 0.55$, substantially 
larger than that observed on cluster scales. Of more interest is a 
region at the upper left of Figs. 6 and 11, where 
$w(t_i) \lesssim 1.5$ and $0.9 \lesssim \ln \beta \lesssim 1.8$. This 
region is allowed by the SNe data, yields $H_0 t_0 \sim 0.9 \pm 0.1$, 
and corresponds to $\Omega_{m0} \simeq 0.3 - 0.4$.

\begin{figure} 
\hspace*{1.3in} 
\psfig{file=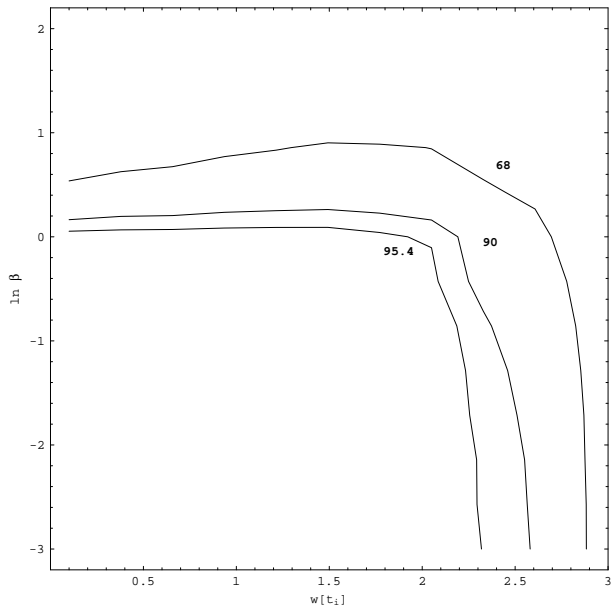,height=8.cm,width=8.cm} 
\vspace*{0.0in} 
\caption{SNe Ia constraints on the parameter space of exponential 
potential models.} 
\end{figure}

\section{Conclusions}

We have studied the observational implications  
of cosmological models in which a 
classical scalar field dominates the energy density of the Universe 
at recent epochs. The motivation for introducing these fields was to provide 
a dynamical model for the cosmological constant, which is favored 
by observations but whose origin remains obscure. These three 
examples were chosen from the literature in order to illustrate the 
range of expected behavior in scalar field models. To date, the 
most stringent observational constraint on the cosmological constant 
comes from recent observations of distant Type Ia supernovae, 
$\Omega_\Lambda < 0.51$ at $2-\sigma$. We 
have extended this constraint to the scalar field ``dynamical'' $\Lambda$ 
models. Since the effective equation of state of an evolving scalar field 
differs from that of a conventional cosmological constant, there 
are regions of parameter space for which the model predictions 
are consistent with the SNe observations, even at relatively high 
values of $\Omega_\phi$. In particular, there are viable 
scalar field models with $\Omega_{m0} \simeq 0.2-0.3$, consistent 
with cluster observations, and which are spatially flat, consistent 
with the predictions of inflation. We close by stressing that 
the high-redshift SNe results are preliminary, based on a new 
technique applied to a very small sample. The on-going SNe searches 
are continuing to discover SNe Ia; as the sample grows and 
the systematic effects become better studied, the  
constraints on cosmological parameters, and on 
the kinds of cosmological models studied here, should become more 
robust. 
 
\acknowledgments

We thank Saul Perlmutter for helpful discussions. 
This research was supported in part by 
the DOE and NASA grant NAG5-2788 at Fermilab and by the Brazilian 
agency CNPq.


\begin{references}
\bibitem{carlberg}R. Carlberg, {et al.}, Astrophys. J. {\bf 462}, 32
(1996).
\bibitem{dbw97}A. Dekel, D. Burstein, and S. White, in {\it Critical 
Dialogues in Cosmology}, ed. N. Turok (World Scientific, 1997),  
astro-ph/9611108.
\bibitem{hubble}W. Freedman, in {\it Proc. of the 18th Texas Symposium 
on Relativistic Astrophysics}, eds. A. Olinto, J. Frieman, and D. Schramm 
(World Scientific, in press), astro-ph/9706072. 
\bibitem{gc}B. Chaboyer, P. Demarque, P. J. Kernan, and L. M. Krauss,
Science {\bf 271}, 957 (1996).
\bibitem{gc2}B. Chaboyer, P. Demarque, P. J. Kernan, and L. M. Krauss, 
astro-ph/9706128. 
\bibitem{lcdm}G. Efstathiou, S. Maddox, and W. Sutherland, Nature
(London) {\bf 348}, 705 (1990); L. Kofman, N. Gnedin, and N. Bahcall,
Astrophys. J. {\bf 413}, 1 (1993).
\bibitem{pd}J. A. Peacock and S. J. Dodds, Mon. Not. R. Astron. Soc. {\bf 267},
1020 (1994).
\bibitem{kochanek}C. S. Kochanek, Ap. J. {\bf 466}, 638 (1996) and 
references therein. This limit is, however, subject to 
possible systematic errors, including extinction by dust in the lensing
galaxy. See, e.g., S. Malhotra, J. E. Rhoads, and E. L. Turner, 
Mon. Not. Roy. Astron. Soc., in press (1997), astro-ph/9610233.
\bibitem{primack} A. Klypin, J. Primack, and J. Holtzman, Ap. J. {\bf 466}, 
13 (1996).
\bibitem{dynlam}M. Ozer and M. O. Taha, Nucl. Phys. {\bf B287}, 776 (1987); 
K. Freese, F. C. Adams, J. A. Frieman, and E. Mottola, Nucl. Phys. 
{\bf 287}, 797 (1987); W. Chen and Y. S. Wu, Phys. Rev. {\bf D41}, 695 (1990); 
J. C. Carvalho, J. A. S. Lima, and I. Waga, Phys. Rev. {\bf D46}, 2404 (1992); 
V. Silveira and I. Waga, Phys. Rev. {\bf D50}, 4890 (1994).
\bibitem{PR}P. J. E. Peebles and B. Ratra, Ap. J., {\bf 325}, L17 (1988).
\bibitem{RP}B. Ratra and P. J. E. Peebles, Phys. Rev. {\bf D37}, 3406 (1988).
\bibitem{fhsw}J. A. Frieman, C. T. Hill, A. Stebbins, and I. Waga, 
Phys. Rev. Lett. {\bf 75}, 2077 (1995).
\bibitem{RQ}B. Ratra and A. Quillen, Mon. Not. R. Astron. Soc. {\bf 259}, 
738 (1992).
\bibitem{bw}L. F. Bloomfield Torres and I. Waga, Mon. Not. R. Astron. Soc. 
{\bf 279}, 712 (1996).
\bibitem{cdf97}K. Coble, S. Dodelson, and J. A. Frieman, 
Phys. Rev. {\bf D55}, 1851 (1997). 
\bibitem{cds97}R. R. Caldwell, R. Dave, and P. J. Steinhardt, astro-ph/9708069.
\bibitem{vl97}P. T. P. Viana and A. R. Liddle, astro-ph/9708247.
\bibitem{perl96}S. Perlmutter {\it et al.}, Ap. J. {\bf 483}, 565 (1997). 
\bibitem{schmidt97}B. Schmidt, B., etal. in {\it Thermonuclear Supernovae}, 
eds. R. Ruiz-Lapuente, R. Canal, and J. Isern (Dordrecht: Kluwer, 1997).
\bibitem{vs}P. J. Steinhardt, Nature {\bf 382}, 768 (1996); 
M. Turner and M. White, astro-ph/9701138; V. Silveira and I. Waga, 
astro-ph/9703185.
\bibitem{fhw} J. Frieman, C. Hill, and R. Watkins, Phys. Rev. D {\bf 46},
1226 (1992).
\bibitem{fukyan}M. Fukugita and T. Yanagida, preprint YITP/K-1098 (1995).
\bibitem{kr97}W. Kinney and A. Riotto, Fermilab-Pub-97-090-A 
(hep-ph/9704388) and references therein.
\bibitem{lucch}See, e.g., L. F. Lucchin and S. Matarrese, Phys. 
Rev. {\bf D32}, 217, (1985); J. J. Halliwell, Phys. Lett. {\bf B 185}, 341, 
(1987); A. B. Burd and J. D. Barrow, Nucl. Phys {\bf B308}, 929 (1988);
A. R. Liddle, Phys. Lett. {\bf B 220}, 502 (1989).
\bibitem{sahni92}V. Sahni, H. Feldman, and A. Stebbins, Ap. J. {\bf 385}, 1 
(1992).
\bibitem{hamuy}M. Hamuy {\it et al.}, Astron. J. {\bf 109}, 1 (1995). 
M. Hamuy {\it et al.}, Astron. J. {\bf 112}, 2391 (1996).
\bibitem{phil93}M. M. Phillips, Ap. J. {\bf 413}, L105 (1993). 
\bibitem{rpk95}A. G. Riess, W. H. Press, and R. P. 
Kirshner, Ap. J. {\bf 438}, L17 (1995).
\bibitem{rpk96}A. G. Riess, W. H. Press, and R. P. Kirshner, Ap. J. {\bf 473}, 
88 (1996).

\end{references}
\end{document}